\documentclass[twocolumn,showpacs,preprintnumbers,amsmath,amssymb]{revtex4}

\usepackage{graphicx}
\usepackage{dcolumn}
\usepackage{bm}
\newcommand{\bi}[1]{\mbox{\boldmath$#1$}}
\newcommand{\av}[1]{\langle{#1}\rangle}
\def\be{\begin{equation}}
\def\en{\end{equation}} 
\def\p{\partial }  
\def\ve{\varepsilon}

\def\bphi{\bar{\phi}}

\def\bea{\begin{eqnarray}}
\def\ena{\end{eqnarray}}

\begin{document}
\preprint{APS}
\title{Precipitation  of water 
from aqueous mixtures  with addition of 
hydrophilic ions }

\author{Ryuichi Okamoto}
\author{Akira Onuki}
\affiliation{Department of Physics, Kyoto University, Kyoto 606-8502, Japan}

\date{\today}

\begin{abstract}
We examine phase separation in aqueous  mixtures 
at  fixed amounts  of    hydrophilic  monovalent ions. 
When  water is the  minority component, 
preferential solvation 
can stabilize water  domains enriched with ions.  
 This ion-induced precipitation 
occurs  in  wide ranges of the temperature  
and the average composition where  
the  solvent would be  in one-phase 
states without ions. 
The volume fraction of 
such water domains is decreased  to zero 
as the interaction parameter 
$\chi$ (dependent on the temperature) 
is decreased toward a critical value for each 
 average composition.  
\end{abstract}


\maketitle

%
%

In  fluid mixtures  containing  water, 
phase separation behavior 
can be drastically changed by  a small amount of  a 
salt \cite{polar1}.  
More strikingly,  many  
groups have observed long-lived 
heterogeneities (sometimes extending  over 
a few micrometers)  
in one-phase states \cite{So} and a  third phase 
visible  as a thin plate  
at  a liquid-liquid interface in two-phase states \cite{third}. 
Very recently, mesophases with  lamellar 
or onion structures have been found 
for an antagonistic salt  composed of 
hydrophilic and hydrophobic ions \cite{Sadakane}.  
Dramatic  ion effects  are ubiquitous 
in various soft matters.  
For example, in polyelectrolytes,  the phase behavior 
sensitively depend on  the  
degree of ionization and  the composition 
for mixture   solvents \cite{Bl} 
and large-scale heterogeneities have been 
observed  \cite{Amis}.  In these phenomena, 
the  solvation interaction   among 
ions and polar  molecules should
play a major role together with the Coulomb  
interaction among charges \cite{Is}. 
Recently some theoretical efforts have been made 
to elucidate the solvation effects 
in phase transitions in electrolytes and polyelectrolytes 
 \cite{OnukiPRE,Araki,Okamoto}. 
In this Letter, we consider  hydrophilic 
monovalent ion pairs such as Na$^+$ and Cl$^-$ 
in  a binary mixture  of water and a less polar 
component (oil or alcohol) and 
examine  ion-induced precipitation.

  Neglecting the 
electrostatic interaction but accounting for the 
solvation interaction, we first 
 consider  a  binary mixture in a cell with a  fixed volume $V$. 
Here ions  constitute  the 
 third component with density $n({\bi r})$. 
The volume fractions 
of water,  oil, and ions   are  written as 
$\phi({\bi r})$ and  $\phi'({\bi r})$, and 
$v_I n({\bi r})$, respectively, where $v_I$ is the ionic   
 volume. If the two solvent species  have the same 
molecular volume $v_0$, their densities are 
$\phi/v_0$ and  $\phi'/v_0$.  The 
 space-filling  condition reads $\phi+\phi'+ 
v_I n=1$.  The ion volume fraction 
is assumed to be  small or 
$ 
v_I n\ll 1,
$ 
which is   easily satisfied  
for small ions with $v_I \ll v_0$. In this Letter  we thus set   
$\phi'=1-\phi$, which simplifies the calculations.

When  the ions  have a 
 strong preference of  water over oil, 
we set up the  free energy  as 
\be 
\frac{F}{T} = \int d{\bi r}\bigg[ 
\frac{f(\phi)}{T}  + \frac{C}{2}|\nabla \phi|^2+  
  {n}\ln (nv_0)  - g n\phi\bigg], 
\en 
where $C$ is a positive constant 
and $g$ is a positive parameter representing 
the ion  preference  of  water. 
The space integral is within the cell.  
The Boltzmann constant will be set equal to unity 
and then the temperature $T$ represents the thermal energy  of a 
particle.  The free energy density  $f(\phi) $ is taken to be     
the simple form,  
\be 
v_0 f/T  = 
 {\phi}  \ln\phi + (1-\phi)\ln (1-\phi) 
+ \chi \phi (1-\phi) , 
\en 
where $\chi$ is the interaction parameter 
 dependent  on  $T$ 
and its  mean-field critical value is $2$   without  ions. 
 We  fix  the total particle numbers 
of the three components as 
\be 
{\bar n}= \int d{\bi r} n/V, 
\quad \bphi = \int d{\bi r} \phi/V,
\en 
In equilibrium   the homogeneity 
of the ion  chemical potential $\delta F/\delta n$ yields  
\be 
n= {\bar n} e^{g\phi}/\av{e^{g\phi}},
\en 
where $\av{e^{g\phi}}= \int d{\bi r}e^{g\phi}/V$ is 
the space average of $e^{g\phi}$. 
Substitution of Eq.(4) into $F$ in Eq.(1) gives  
\be 
\frac{F}{T} = \int d{\bi r}\bigg[ 
\frac{f(\phi)}{T}  + \frac{C}{2}|\nabla \phi|^2\bigg]
+  V\bar{n}\ln[ \bar{n}v_0/\av{e^{g\phi}}].  
\en

In equilibrium the  chemical potential difference 
  $h= \delta F/\delta \phi$  
for the composition is also homogeneous. Here,  
\be 
h= f'(\phi)  -TC\nabla^2 \phi -Tg n 
\en 
where $f'= \partial f/\partial \phi$. 
 Around a planar interface 
varying along the $z$ axis, we obtain 
$TC (d\phi/dz)^2=2H(\phi)$, where  
\be 
H (\phi) = f(\phi)-f(\phi_\alpha)-T(n-n_\alpha) 
 -h(\phi-\phi_\alpha).    
\en  
We  suppose  coexistence of 
a water-rich  phase $\alpha$ 
and an oil-rich phase $\beta$ 
with $\Delta\phi=\phi_\alpha-\phi_\beta>0$. 
The compositions and the ion  densities 
in the two phases are written as 
$\phi_\alpha$,  $\phi_\beta$, $n_\alpha$, and $n_\beta$, 
respectively. The volume fraction of the 
 phase $\alpha$  is  
denoted by $\gamma_\alpha$. Since that of the phase $\beta$  is 
 $\gamma_\beta=1-\gamma_\alpha$, we have 
\be 
\gamma_\alpha= ({{\bar \phi}-\phi_\beta})/{\Delta \phi}
=  ({{\bar n}-n_\beta})/{\Delta n}, 
\en  
where $\Delta n=n_\alpha-n_\beta$. 
From Eq.(4) the  ratio of the bulk 
 ion  densities  is written as 
$n_\alpha/n_\beta= \exp({g\Delta\phi})\gg 1$ from Eq.(4), 
where we  assume $g\Delta\phi\gg 1$. 
Neglecting 
the surface free energy,  we express  $F$ as 
\be 
\frac{F}{V}= \gamma_\alpha f_\alpha 
+ \gamma_\beta f_\beta - T{\bar n} \log [ 
(\gamma_\alpha e^{g\phi_\alpha}+ \gamma_\beta
e^{g\phi_\beta})/{\bar n}v_0],
\en   
where $f_\alpha=f(\phi_\alpha)$ and $f_\beta =f(\phi_\beta)$.
We minimize this  $F$ with respect to 
$\gamma_\alpha$, $\phi_\alpha$, and $\phi_\beta$ at fixed $\bphi$ 
to obtain 
\bea 
&&h= f_\alpha'-Tgn_\alpha=f_\beta'-Tgn_\beta,\\  
&&{f_\alpha-f_\beta}- T\Delta n=h{\Delta\phi}, 
\ena 
where $f_\alpha'=f'(\phi_\alpha)$ and $f_\beta'=f'(\phi_\beta)$. 
These equations also follow from  Eqs.(6) and (7).

\begin{figure}[th]
\begin{center}
\includegraphics[scale=0.49]{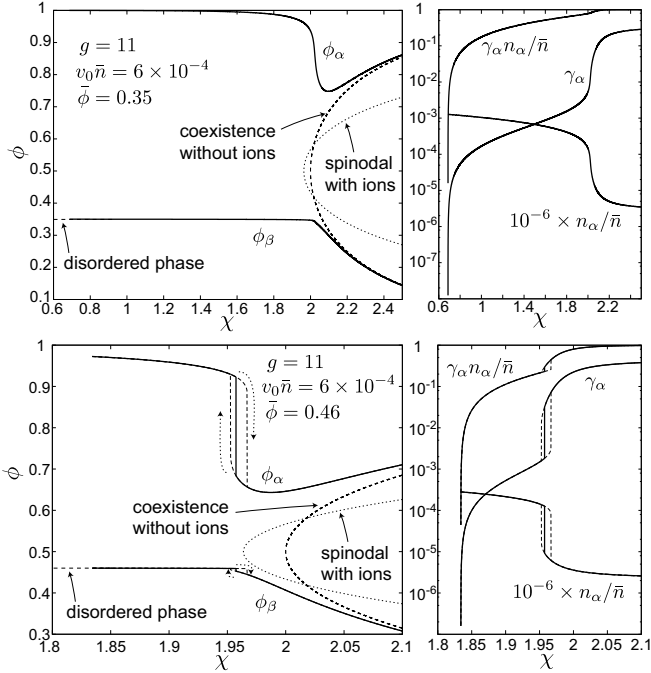}
\caption{ 
Compositions $\phi_\alpha$ and $\phi_\beta$ (left) 
and $\gamma_\alpha$, $\gamma_\alpha n_\alpha/{\bar n}$, 
and $n_\alpha/{\bar n}$ (right) vs $\chi$,  
where ${\bar n}= 6\times 10^{-4}v_0^{-1}$  and $g=11$. 
For $\bar{\phi}= 0.35$ (top), $\phi_\alpha$ 
continuously changes for $\chi> \chi_{\rm p} =0.687$. 
For $\bar{\phi}= 0.46$ (bottom),  
$\phi_\alpha$ jumps at $\chi \cong 2$, where 
$\chi_{\rm p}=1.834$. 
Shown also are the coexistence curve without ions 
and the spinodal curve with ions. 
The latter follows 
from a shift of the spinodal curve without ions 
 by $g^2v_0{\bar n}/2$ to the left \cite{OnukiPRE}.}
\end{center}
\end{figure}

\begin{figure}[th]
\begin{center}
\includegraphics[scale=0.5]{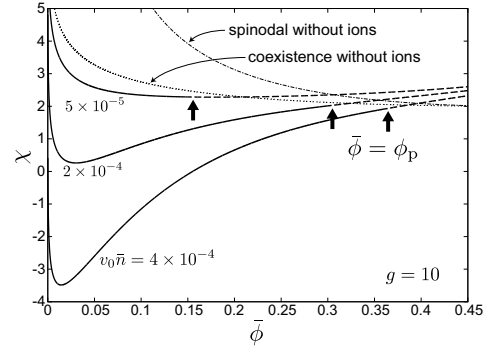}
\caption{$\chi=\chi_{\rm p}(\bphi)$ 
for  $v_0 \bar{n}= 4\times 10^{-4}$, 
$2\times 10^{-4}$, and $5\times 10^{-5}$. 
Precipitation occurs  for 
$\bphi<\phi_{\rm p}$ (arrows). 
Shown also are the coexistence 
and  spinodal curves  without  ions.}
\end{center}
\end{figure}

Our main finding is that 
there appears a precipitation branch of 
$\phi=\phi_\alpha \cong 1$  
in the region $\chi<2$ for $g\gg 1$. 
In  Eq.(10) we set $f'_\alpha \cong v_0^{-1}
 T [-\log (1-\phi_\alpha)-\chi]$ to obtain   
\be 
{1-\phi_\alpha}\cong A_\alpha \exp[-g G(\phi_\beta)],
\en 
where $A_\alpha=\exp(\chi + v_0 f'_\beta/T)$. Here we introduce 
\bea 
G(\phi) &=& -v_0[f(\phi) + (1-\phi)f'(\phi)]/T\nonumber\\
&=& -\log\phi- \chi(1-\phi)^2, 
\ena 
where the  second line follows from Eq.(2). 
By setting $f_\alpha\cong 0$ in Eq.(11) 
we also find $v_0 n_\alpha= G(\phi_\beta)$. 
Outside the spinodal curve without ions, 
we notice $dG/d\phi= -v_0 (1-\phi) \p^2f/\p \phi^2<0$, 
which leads  to  $G(\phi)>0$  from $G(1)=0$. 
Thus the right hand side 
of Eq.(12) is small or $\phi_\alpha \cong 1$ for large $g$.   
The equation for $\phi_\beta$ is given by   
\be 
v_0n_\alpha= 
G(\phi_\beta)= v_0{\bar n} 
\Delta\phi/[\bar{\phi}-\phi_\beta+  e^{-g\Delta\phi}\Delta\phi]
\en 
where  we may set $\Delta\phi\cong 1-\phi_\beta$. 
As $\gamma_\alpha= ({\bar\phi}-\phi_\beta)/\Delta\phi$ 
decreases to zero, $\phi_\beta$ and $n_\alpha$ 
increase  up to   
$\bphi$ and  $v_0 {\bar n}e^{g(1-\bphi)}$, 
respectively. Remakably, $\phi_\alpha$ and $\phi_\beta$ 
depend on $\bphi$. 
From the second line of Eq.(13), 
the precipitation  branch exists only 
for $\chi>\chi_{\rm p}({\bar \phi})$. The 
 lower bound  is 
\be 
\chi_{\rm p}= 
 [-\log (1-\bphi)- v_0{\bar n}e^{g(1-\bphi)}]/(1-\bphi)^2,  
\en 
where 
$\gamma_\alpha \to 0$ as $\chi \to \chi_{\rm p}$.

To easily understand   the mathematics, 
let us focus on  the case 
 $\gamma_\alpha \ll 1$, where 
$f_\alpha \cong 0 $ and $f_\beta  \cong 
f(\bphi) - f'_\beta 
(1-\bphi)\gamma_\alpha $ in Eq.(9). 
Treating $\gamma_\alpha$ as an order parameter, 
we write the deviation 
$\Delta F=F(\gamma_\alpha)-F(0)$ 
of the free energy  from the one-phase  value 
$F(0)$  as 
\be 
{\Delta F}/T{V}\cong   A_1 \gamma_\alpha - {\bar n} \log ( 
1+ B_1 \gamma_\alpha ) ,
\en   
where $A_1=v_0^{-1} G(\bphi)+\bar{n}g (1-\bphi)$  
and $B_1= e^{g\Delta\phi}-1\cong  e^{g\Delta\phi}$. 
For $w \equiv {\bar n}B_1/A_1>1$, 
 $\Delta F$ has a negative minimum given by 
$-TVA_1 (w\log w-w+1)/B_1<0$ attained at 
\be 
\gamma_\alpha=(w-1)/B_1\cong 
 v_0 {\bar n}/G({\bar \phi}) -e^{-g(1-\bphi)}, 
\en 
which  is consistent with Eq.(14) in the limit 
 $\phi_\beta\to \bphi$. 
The condition  $w>1$ is 
equivalent to $\chi>\chi_{\rm p}$ for $g\gg 1$.

Figure 1 gives 
 the phase diagrams in the $\phi$-$\chi$ plane 
with  ${\bar n}= 6\times 10^{-4}v_0^{-1}$  and $g=11$. 
In the first case of 
$\bphi=0.35$,  $\phi_\alpha$ changes 
continuously  and  is minimum at $\chi=2.05$, 
where   
$\chi_{\rm p}=0.687$ and the maximum of $n_\alpha$ is 
$0.381v_0^{-1}$ at $\chi=\chi_{\rm p}$. 
In the second case of $\bphi=0.46$, 
where $\chi_{\rm p}=1.834$ and $n_\alpha=0.0842v_0^{-1}$ at 
$\chi= \chi_{\rm p}$, we find that 
$\phi_\alpha$ changes discontinuously along a hysteresis 
loop in the range $1.953<\chi<1.967$. 
 In  equilibrium,   $F$ is   minimized and 
the resultant   discontinuous  transition  is  at 
$\chi=1.957$.  In Fig.2,  we  display  curves of 
$\chi=\chi_{\rm p}(\bphi)$ 
for three values of $\bar n$ with  $g=10$. Each curve   
assumes a minimum at small $\bphi$ 
far away from the  coexistence curve 
without ions.   With increasing $\bar\phi$, the precipitation 
branch shrinks and disappears as  $\bphi \to \phi_{\rm p}$, 
where $\phi_{\rm p}$ is a critical composition less than $0.5$. 
As  functions of $\bphi$ and 
$\chi$, we show $\gamma_\alpha$ in Fig.3 
in the continuous case of $g=10$ 
and  $\phi_\alpha$ in Fig.4  
in the discontinuous case of $g=11$.

Without the electrostatic interaction, 
the surface tension of our system 
is expressed as the integral $
\sigma= \int dz CT(d\phi/dz)^2$ 
around an  interface varying along the $z$ axis \cite{OnukiPRE}.  
 Use of Eq.(7) gives 
\be 
\sigma= 
(2CT)^{1/2} \int_{\phi_\beta}^{\phi_\alpha}
 d\phi H(\phi)^{1/2},
\en    
where $C$ is assumed to be a constant. 
In Fig.5, we display the function $[2H(\phi)v_0/T]^{1/2}$ 
for $g=11$.  Here  $H(\phi) \cong f''(\phi_\beta) 
(\phi-\phi_\beta)^2/2$ as $\phi \to \phi_\beta$ 
with $f''= d^2f/d\phi^2$. Thus we 
obtain   $\sigma \sim TC (\Delta\phi)^2/2\xi$, 
where $\xi = (f''/CT)^{-1/2}$ is the correlation length 
at  $\phi= \phi_\beta \cong 
\bphi$.

\begin{figure}[th]
\begin{center}
\includegraphics[scale=0.75]{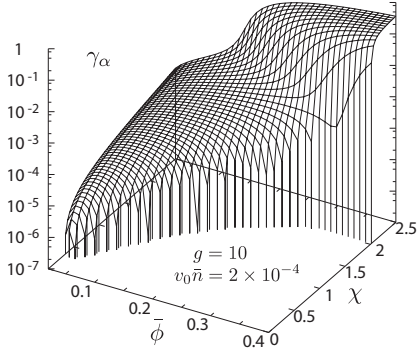}
\caption{Volume fraction of 
the water-rich phase $\gamma_\alpha$ as 
a function of $\chi$ and $\bphi$. 
for ${\bar n}= 2\times 10^{-4}v_0^{-1}$  and $g=10$. 
In  this case  $\gamma_\alpha$ decreases continuously to zero 
as $\chi \to \chi_{\rm p}$.  
}
\end{center}
\end{figure}

\begin{figure}[th]
\begin{center}
\includegraphics[scale=0.75]{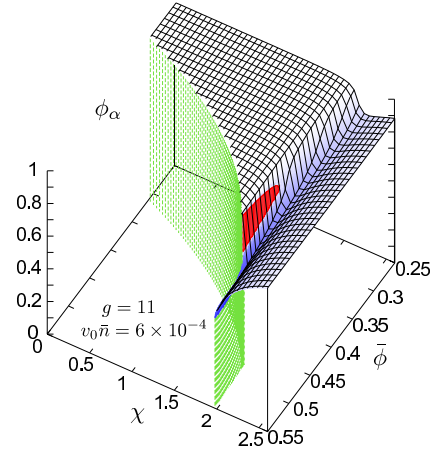}
\caption{(Color on line) Composition of the water-rich phase 
$\phi_\alpha$  
for  $v_0 \bar{n}= 6 \times 10^{-4}$ 
and $g=11$. In this case  
a discontinuous transition 
occurs for $\chi\cong 1.95$ and ${\bar \phi}>0.395$ 
 (in red). 
One-phase states are realized 
for  $\chi<\chi_{\rm p}$ 
or for ${\bar \phi}>\phi_{\rm p}=0.473$ 
and $\chi<2$ (in white).  
 For $\chi>2$, $\phi_\alpha$ is little affected 
by ions.    
}
\end{center}
\end{figure}

\begin{figure}[th]
\begin{center}
\includegraphics[scale=0.5]{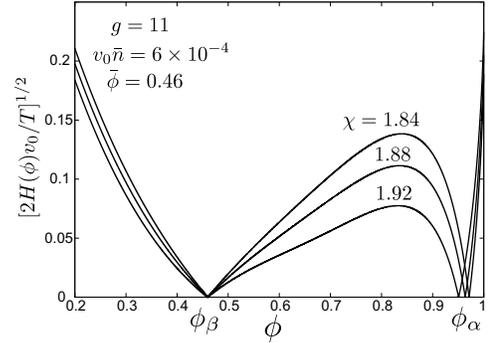}
\caption{$[2H(\phi)v_0/T]^{1/2}$ 
vs $\phi$,  vanishing  at $\phi=\phi_\alpha$ 
and  $\phi_{\beta}$, where  
$\bphi=0.46$, ${\bar n}= 6\times 10^{-4}v_0^{-1}$,
  and $g=11$.}
\end{center}
\end{figure}

\begin{figure}[th]
\begin{center}
\includegraphics[scale=0.44]{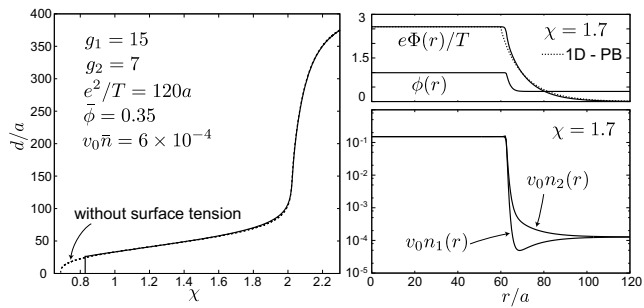}
\caption{Left: 
Numerical droplet radius $d/a$ vs $\chi$ (bold line) 
together with 
the  theoretical curve without the surface 
free energy (dotted line).  
Right: Normalized potential $e\Phi(r)/T$ and water volume fraction 
$\phi(r)$ (top), and normalized ion densities 
$v_0n_1(r)$ and $v_0n_2(r)$ (bottom), where  $\chi=1.7$. Here 
 $g_1=15$, $g_2=7$, ${\bar \phi}=0.35$, 
and  $v_0{\bar n}=6\times 10^{-4}$.  
}
\end{center}
\end{figure}

Including the electrostatic interaction, we 
next consider aqueous  mixtures 
containing a   hydrophilic monovalent 
salt. The cation and anion densities are written as 
$n_1$ and $n_2$, whose  total amounts are fixed 
as 
\be 
\int d{\bi r} n_1= \int d{\bi r}n_2=V{\bar n}/2.
\en 
The  electric potential $\Phi$ 
satisfies the Poisson equation 
$\nabla \cdot \ve \nabla\Phi= -4\pi e
(n_1-n_2)$, where the dielectric constant 
$\ve(\phi)$ can depend on $\phi$.  
The  free energy $F$  reads \cite{OnukiPRE}
\bea 
\frac{F}{T} &=& \int d{\bi r}\bigg[ 
\frac{f(\phi)}{T}  + \frac{C}{2}|\nabla \phi|^2+ 
\frac{\ve|{\nabla \Phi}|^2}{8\pi T}
\nonumber\\ 
&+&  \sum_{i=1,2} [{n_i}\ln (n_i v_0)  - g_in_i \phi] \bigg ]. 
\ena 
The   ion   chemical potentials    
due to 	solvation, written as   $\mu_{\rm sol}^i(\phi)$,     
strongly depend on $\phi$. 
The interaction  terms ($\propto g_i$) in $F$  follow   for  
 the linear forms  $\mu_{\rm sol}^i(\phi)= 
\mu_{0}^i- Tg_{i}\phi$   
(where the first terms  are  irrelevant constants). 
This  linear dependence  is adopted to 
gain the physical consequences in the simplest 
manner.  For each ion species $i$, 
 the solvation-chemical-potential difference 
between the two phases is given by 
 $\Delta\mu_{\alpha\beta}^{i}=
 Tg_i\Delta\phi$, 
which is  the Gibbs transfer 
free energy  in electrochemistry 
\cite{Hung}.  In aqueous solutions,   
 $g_{i}\gg 1$   for hydrophilic small  ions, 
while $g_{i}<0$ for hydrophobic ions \cite{Hung,OnukiPRE}.  
We minimize $F$ with respect to $n_i$ under Eq.(19) to obtain 
\be
n_i= n_i^0 \exp [g_i\phi \mp  e\Phi/T], 
\en  
where $-$ is for $i=1$, $+$ is for $i=2$, and  
$n_i^0=  {\bar n}/[2\int d{\bi r}
\exp (g_i\phi \mp  e\Phi/T)]$. The composition profile is determined by 
the homogeneity of $\delta F/\delta \phi$. 
Here we neglect the image interaction,  
whose role is reduced compared to that of the solvation 
interaction for not small ion densities 
\cite{OnukiPRE,Levin}.

For  $g_1=g_2=g$, we have $n_1=n_2=n/2$ and $\Phi=0$,  
so  $F$ in Eq.(20) reduces to $F$ in Eq.(1). For 
$g_1\neq g_2$,  an electric double layer appears 
at the interface with  a  potential difference 
$\Delta\Phi = T(g_1-g_2) \Delta\phi/2e$ across it, 
but the bulk phase relations (10) and (11) 
still hold with 
\be 
g=(g_1+g_2)/2.
\en  
In Fig.6, we give numerical results 
for  $g_1=15$, $g_2=7$, ${\bar \phi}=0.35$, 
and  $v_0{\bar n}=6\times 10^{-4}$. 
The space unit is $a=v_0^{1/3}$. 
We suppose  a spherical water-rich droplet 
with radius $d$ placed at the center 
of a spherical cell with radius $R=600a$. 
Then $\gamma_\alpha=(d/R)^3$. 
The dielectric constant is of the form 
 $\ve=40 (1+\phi)$. We also set  
 $aC= \chi$ and $e^2/ T=120a$. In the left,  
the droplet disappears  at $d =26.0a$, where 
$w$ in Eq.(17) is 1.08. This critical radius 
follows   if the bulk free energy 
$\Delta F$ in Eq.(9) is equated with  the minums of the surface   
free energy  $4\pi \sigma d^2$. 
In the right, we set $\chi=1.7$ to display  the profiles of 
$\Phi(r)$, $\phi(r)$, $n_1(r)$, and $n_2(r)$. 
We obtain $\phi_\alpha 
=0.993$ and  $n_{\alpha}
= 0.352v_0^{-1}$ within the droplet  and  
 $\phi_\beta=0.349$ and  
$n_{\beta}=2.55\times 10^{-4}  v_0^{-1}$ 
outside it. In Fig.6, the potential $\Phi(r)$ 
relaxes with the  Debye length  
$\kappa_\beta^{-1}=  11.8 a$ and is well  
fitted to  the one-dimensional 
solution of the nonlinear Poisson-Boltzmann 
 equation (dotted line)  
with $\Delta \Phi=3T/e$ \cite{OnukiPRE}. Here  
$n_\alpha$ and $n_\beta$ 
are the bulk values of $n=n_1+n_2$.

In future we  should  explain 
the experimental findings of large-scale 
heterogeneities \cite{So}. We  note that 
one-phase  states are 
 metastable outside  the spinodal curve with ions 
($\chi< 1/2\bphi(1-\bphi)-g^2v_0{\bar n}/2$)  in Fig.1. 
Thus precipitation from a one-phase state 
should be triggered by some impurities 
and/or  hydrophilic  walls.  
 We  also note that the 
wetting transition of  aqueous mixtures 
is much influenced  by the ion-induced 
precipitation mechanism.

 Experiments are informative, 
where  the temperature, the water volume fraction, 
and the  salt amount are  varied.  
We mention an  experiment by 
Leunissen  {\it et al.} \cite{Bla}, 
where micron-sized water droplets containing ions 
formed  a crystal  
in an  oil   with low dielectric constant ($\ve_{\rm oil}=4-10$) 
 without a surfactant. Graaf {\it et al.} \cite{Roij} 
ascribed its origin to the screened Coulomb interaction 
among droplets.  We also   propose experiments of the 
 salting-out effect of polyelectrolytes  
in water-alcohol \cite{Bl}, where the degree of  
ionization much increases with accumulation of water 
around the polymers
 \cite{Okamoto}.

\begin{acknowledgments}
This work was supported by Grant-in-Aid 
for Scientific Research on Priority 
Area ``Soft Matter Physics'' from the Ministry of Education, 
Culture, Sports, Science and Technology of Japan.
\end{acknowledgments}

\end{document}